\documentclass{aa}
\usepackage{graphics,epsfig,amsmath,amssymb,amstext}

\begin{document}

\title{GRBs and the 511~keV emission of the Galactic bulge}

\author{E. Parizot\inst{1}, M. Cass\'e\inst{2,3}, R. Lehoucq\inst{2} \& J. Paul\inst{2,4}}

\offprints{parizot@ipno.in2p3.fr}

\institute{Institut de Physique Nucl\'eaire d'Orsay, IN2P3-CNRS/Universit\'e Paris-Sud, 91406 Orsay Cedex, France \and DAPNIA/Service d'astrophysique, CEA-Saclay, 91191 Gif-sur-Yvette, France \and Institut d'Astrophysique de Paris, 98 bis, bd Arago, 75014 Paris, France\and F\'ed\'eration de Recherche, Astroparticule et Cosmologie, Univ. Paris 7, Coll\`ege de France, 75231 Paris Cedex 05}

\date{Received date; accepted date}

\abstract{We consider the phenomenology of the 511~keV emission in the Galactic bulge, as recently observed by INTEGRAL, and propose a model is which the positrons are produced by gamma-ray bursts (GRB) associated with mini starbursts in the central molecular zone (CMZ). We show that the positrons can easily diffuse across the bulge on timescales of $\sim 10^7$~yr, and that their injection rate by GRBs is compatible with the observed fluxes if the mean time between two GRBs in the bulge is $\sim 8\,10^4$~yr$\times (E_{\mathrm{GRB}}/10^{51}\,\mathrm{erg})$. We also explain the low disk-to-bulge emission ratio by noting that positrons from GRBs in the Galactic disk should annihilate on timescales of $\la 10^4$~yr in the dense shell of the underlying supernova remnant, after the radiative transition, while the remnants of GRBs occurring in the hot, low-density medium produced by recurrent starbursts in the CMZ become subsonic before they can form a radiative shell, allowing the positrons to escape and fill the whole Galactic bulge. If the mean time between GRBs is smaller than $\sim 10^4\,\mathrm{E}_{51}$~yr, INTEGRAL should be able to detect the (localized) 511~keV emission associated with one or a few GRB explosions in the disk.

}

\maketitle

\section{Introduction}
\label{sec:introduction}

The region around the Galactic Center (GC) has been known for almost three decades to be an intense source of e$^+$e$^-$ annihilation emission (Paul 2004). Recent observations with the gamma-ray satellite INTEGRAL confirmed the integrated flux level reported by previous experiments, namely $\sim 10^{-3}\,\mathrm{ph\,cm}^{-2}\mathrm{s}^{-1}$, and provided improved spectral resolution of the 511~keV line as well as important constraints on the angular size of the emitting region (Weidenspointner, 2004). The available data are not compatible with a single point source and are well modeled by a spherically symmetric gaussian emission profile centered on the GC, with FWHM between 6$^\circ$ and 11$^\circ$, corresponding to a radius between 500 and 800~pc, as would result if positrons were filling a large fraction of the Galactic bulge. While a model including four point sources or more is still compatible with the data, no evidence of such localized emission has been found. Therefore, waiting for improved constraints on the emission map, we shall assume that the data interpret as a diffuse e$^{+}$e$^{-}$ annihilation radiation from positrons distributed all over the Galactic bulge.

The phenomenology of the 511~keV emission comprises two \emph{a priori} distinct aspects: i) the source of the positrons responsible for such an intense radiation, corresponding to $\sim 1.3\,10^{43}$ annihilations per second, and ii) their spatial distribution at annihilation. In a recent study, Boehm et al. (2004) assume that positrons in the Galactic bulge diffuse with a very low diffusion coefficient and essentially remain where they are produced, so that the emission map directly relates to the source distribution. Positrons produced throughout the Galactic bulge by the annihilation of some unknown low-mass dark matter particles could then in principle explain the data (Boehm et al., 2004). Most recently, Bertone et al. (2004) considered gamma-ray bursts (GRBs) as the possible source of the indirectly observed positrons. Since they also assume Bohm-like diffusion, they have to distribute the GRBs all over the bulge. This does not seem reasonable, however, because the bulge is essentially devoid of massive stars which are believed to be associated with the GRBs (Zhang and M{\' e}sz{\' a}ros, 2004).

In this paper, we consider positron transport from a more general point of  view and propose a scenario where positrons are injected episodically by GRBs in the innermost region of the Galaxy, and diffuse throughout the bulge before they annihilate in the ambient, low-density gas. We also address an apparent paradox which is an important part of the phenomenology of the Galactic 511~keV diffuse emission: if the production of positrons is somehow related to the activity of massive stars (via $\beta^+$ nuclei produced by explosive nucleosynthesis episodes or ejected by stellar winds, Schanne et al., 2004 and references therein, or in association with GRBs), then why do we see the 511~keV line from the Galactic bulge and not from the disk, or from the molecular ring where many massive stars concentrate, between 4 and 5~kpc from the GC, or from the very active and nearby Cygnus region, which has also been surveyed by INTEGRAL? Likewise, if the positrons are produced by cosmic-ray interactions in the interstellar medium (ISM), the emission should in principle be distributed throughout the disk with a possible increased flux around superbubbles and the main nearby stellar nurseries. In the model proposed here, this paradox finds a natural solution related to the different ISM environments around GRBs in the Galactic center and in the disk, without additional assumptions about source distribution or particle transport.

\section{Diffuse emission and positron transport}
\label{sec:transport}

\subsection{Relevant timescales}
\label{sec:timescales}

The spatial distribution of the positrons when they annihilate depends on their transport properties as well as their lifetime in the Galactic bulge environment. In the case of positrons injected with energies lower than $\sim 100$~MeV, the magnetic field and the photon field in the bulge are too low for synchrotron and inverse Compton energy losses to dominate. Bremsstrahlung is also negligible and the main energy loss process is associated with Coulombian interactions in the ambient plasma. The corresponding energy loss time, $\tau_{\mathrm{loss}}$, depends on the positron energy, the gas density, $n_{0}$, and its degree of ionization. For positrons between 1 and 100~MeV, one finds $\tau_{\mathrm{loss}}\sim 10^5\, n_{0}^{-1}\mathrm{yr}$, within a factor of 2 or so, where $n_{0}$ is in cm$^{-3}$. The gas density in the bulge is poorly known, although probably much lower than in the disk. We shall assume here $n_{\mathrm{bulge}}\sim 10^{-2}\,\mathrm{cm}^{-3}$, so that $\tau_{\mathrm{loss}}\sim 10^7$~yr.

When the positrons reach energies of a few tens of eV or below, they can either annihilate directly with ambient electrons (free or inside H atoms) or form a positronium by charge exchange or radiative capture. The annihilation timescale depends on the gas temperature and density, and can have very different values from $\tau_{\mathrm{ann}}\sim 10^3$~yr in molecular clouds to $\tau_{\mathrm{ann}}\sim 10^8$~yr in the hot ISM,  with values around $10^5$~yr in the typical warm ISM where most positrons seem to annihilate (as indicated by the positronium fraction; Paul, 2004). For the problem considered here, this timescale is not crucial, because the positrons having reached such low velocities should be more or less tight to the ambient plasma. The distribution of positrons in the bulge is thus expected to be roughly the same as that achieved after a few $\tau_{\mathrm{loss}}$. The only important point is that $\tau_{\mathrm{loss}} + \tau_{\mathrm{ann}} \ge \Delta t_{\mathrm{inj}}$, where $\Delta t_{\mathrm{inj}}$ is the timescale between events injecting positrons in the bulge, so that a steady state can settle. This condition is already ensured by the long energy loss time, independently of $\tau_{\mathrm{ann}}$.

\subsection{Diffusion coefficients}

As recalled in the Introduction, positron transport in the bulge can be reduced to a minimum and even totally ignored by assuming Bohm diffusion, where the typical mean free path of the particles is comparable to their gyroradius in the underlying regular (or long wavelength) magnetic field, $r_{\mathrm{g}} = p/qB \simeq 3.3\times 10^{9}\,E_{\mathrm{MeV}}B_{\mu\mathrm{G}}^{-1}\,\mathrm{cm}$. The diffusion coefficient therefore reads $D_{\mathrm{Bohm}}\equiv \frac{1}{3}r_{\mathrm{g}}v\simeq 3.3\times 10^{19}\,E_{\mathrm{MeV}}B_{\mu\mathrm{G}}^{-1}\, \mathrm{cm}^2\mathrm{s}^{-1}$ for relativistic positrons, and the distance travelled during $\tau_{\mathrm{loss}}$ is less than a pc, even for initial energies of 100~MeV (cf. Boehm et al., 2004).

It should be stressed, however, that the Bohm diffusion regime is not likely to hold in the Galactic bulge, especially for positrons with gyroradii much smaller than the coherence length of the magnetic field. While the Bohm regime is often postulated in order to ensure efficient particle acceleration at the supernova remnant shock, it is most probably realized there (if at all!) thanks to self-generated magnetic turbulence associated with the shock itself and the locally accelerated cosmic-rays. However, in the standard ISM hosting a typically Kolmogorov-like turbulence, the diffusion coefficient will always be much larger than $D_{\mathrm{B}}$, allowing positrons to travel much larger distances.

According to the quasilinear theory of resonant diffusion in inhomogeneous magnetic fields, the diffusion coefficient for particles with gyroradii much smaller than the maximum wavelength of MHD perturbations can be written as $D(E)\sim D_{\mathrm{B}}(r_{\mathrm{g}}/\lambda_{\mathrm{max}})^{1-\beta}\eta^{-1}$, where $\eta\equiv \delta B/B$ and $\beta$ indicates how the magnetic energy is distributed among wavenumbers: $w(k)\propto k^{-\beta}$ (e.g. Berezinsky et al., 1990). Note also that  $\lambda_{\mathrm{max}}$ comes in via the field normalization at the resonant scale. For a Kolmogorov turbulence spectrum, $\beta = 5/3$ and the quasilinear diffusion coefficient is
\begin{equation}
D_{\mathrm{ql}}(E)\simeq 3\,10^{27}\,\mathrm{cm}^2\mathrm{s}^{-1}\times E_{\mathrm{MeV}}^{1/3}B_{\mu\mathrm{G}}^{-1/3}\eta^{-1}\left(\frac{\lambda_{\mathrm{max}}}{1\,\mathrm{kpc}}\right)^{2/3},
\label{eq:DQuasiLinear}
\end{equation}
which is many orders of magnitude larger than the Bohm diffusion coefficient, even if $\eta\simeq 1$, as expected during the periods of activity in the CMZ. In principle, the above theory only applies in the case of a one-dimensional turbulence with a low level of turbulence ($\delta B/B \ll 1$). However, it provides a sufficiently reliable estimate for the order of magnitude calculations of the present paper. Besides, numerical diffusion experiments show excellent agreement between expected and simulated diffusion coefficients (Casse et al., 2002; Parizot, 2004).

With the above diffusion coefficient, the distance travelled by the positrons during $\tau_{\mathrm{loss}}$ would be
\begin{equation}
\begin{split}
\Delta R_{\mathrm{ql}} & \simeq \sqrt{6D_{\mathrm{ql}}\tau_{\mathrm{loss}}} \simeq 550\,\mathrm{pc}\times \eta^{-\frac{1}{2}} \left(\frac{E}{1\,\mathrm{MeV}}\right)^{\frac{1}{6}}\\
&\times\left(\frac{B}{10\,\mu\mathrm{G}}\right)^{-\frac{1}{6}}
\left(\frac{\lambda_{\mathrm{max}}}{1\,\mathrm{kpc}}\right)^{\frac{1}{3}}
\left(\frac{n_{0}}{10^{-2}\,\mathrm{cm}^{-3}}\right)^{-\frac{1}{2}},
\end{split}
\label{eq:DeltaRQL}
\end{equation}
which is quite comparable to the extension of the 511~keV emission reported by INTEGRAL, even for $\eta\sim 1$.

It is worth noting also that a superbubble-like environment should develop in the Galactic bulge during the mini-starburst phases when most positrons should are produced (see below). Therefore, turbulent transport may be expected to prevail, so that the relevant diffusion coefficient would have a lower limit at a value of order $D_{\mathrm{turb}}\sim u\lambda_{\mathrm{max}}$. Assuming typical (root mean square) turbulent velocities of the order of $u\sim 100$ -- $300\,\mathrm{km}\,\mathrm{s}^{-1}$ over scales of 100 -- 30~pc, respectively, one gets $D_{\mathrm{turb}}\sim 3\,10^{27}\,\mathrm{cm}^2\mathrm{s}^{-1}$, and a corresponding diffusion radius after $\tau_{\mathrm{loss}}$ of the order of:
\begin{equation}
\Delta R_{\mathrm{turb}} \simeq 800\,\mathrm{pc}\times \left(\frac{n_{0}}{10^{-2}
\,\mathrm{cm}^{-3}}\right)^{-\frac{1}{2}},
\label{eq:DeltaRTurb}
\end{equation}
independently of the positron initial energy and magnetic field structure.

We conclude that the typical distances travelled by positrons in the Galactic bulge before they annihilate are in general much larger than under the unrealistic assumption of Bohm diffusion, and in good agreement with the extension of the observed annihilation emission. In other words, it is probable that the positrons were injected by one or a few sources in a much smaller region around the Galactic center, e.g. in the so-called Central Molecular Zone where massive stars form, evolve and explode.

Note that the transport of low-energy positrons in the Galactic bulge may still be different from the above estimate, because resonant diffusion does not apply if the particle's gyroradius is smaller than the minimum wavelength of the magnetic perturbations. Turbulent wavelengths ranging at least from $10^8$ to $10^{20}$~cm have been reported in the nearby ($\la 1$~kpc) ISM (Armstrong et al., 1995), but it is not known whether the situation in the low-density Galactic bulge is similar, or if the turbulence can extend down to even lower wavelengths. If the magnetic field inhomogeneities roughly follow the electron density fluctuations, the relevant minimum scale could be comparable to the \emph{ion inertial length}, $\lambda_{\mathrm{i}}=v_{\mathrm{A}}/\Omega_{\mathrm{i}}$, where $v_{\mathrm{A}}$ is the Alfv\'en speed and $\Omega_{\mathrm{i}}$ the ion cyclotron frequency, or to the \emph{ion Larmor radius}, $r_{i}=v_{\mathrm{th}}/\Omega_{\mathrm{i}}$, whichever is largest (e.g. Spangler and Gwinn, 1990). In this case, positrons would remain resonant down to energies $m_{\mathrm{p}}/m_{\mathrm{e}}\sim 1800$ times larger than the thermal energy, i.e. around 1 or 2~keV if the inferred annihilation temperature, $T\simeq 10^4$~K, is assumed (e.g. Paul, 2004). At such low energies, positrons are not expected to freely escape from the Galactic bulge. In any case, the requirement that the positrons remain in the Galactic bulge before they annihilate is common to all models, and one may simply reverse the argument and claim that, although the transport conditions are not well known (mostly because of uncertainties about the gas density and magnetic field structure), they must be such that the positrons do not escape freely. The above argument indicates that this is indeed reasonable.

\section{The source of Galactic bulge positrons}

\subsection{Bulge-to-disk ratio and intermittency}

As recalled in the Sect.~\ref{sec:introduction}, while the annihilation of putative low-mass dark matter particles could be a source of positrons in the Galactic bulge (Boehm et al., 2004), common astrophysical sources in the ISM are generally associated with cosmic rays and stellar activity. However, it seems that the production of $\beta^+$ elements by type-Ia and type-II supernov\ae~is not large enough to account for the observed fluxes, notably because of the small fraction of positrons which do not annihilate in the dense ejecta and can actually escape in the ISM (Cass\'e et al., 2004). In addition, in all the above cases, most of the sources are found in the Galactic disk rather than in the bulge, and it seems difficult to satisfy the observational constraint that the bulge-to-disk emission ratio is larger than 0.5--0.8 (Weidenspointner et al., 2004). It is worth noting that Cass\'e et al. (2004) have shown that in the case of type~Ic asymmetric explosion, possibly associated with GRBs, the total positron yield could be sufficient to marginally account for the data.

If the positrons are associated with some phenomenon which can take place in the whole Galaxy, a large bulge-to-disk emission ratio can most naturally be obtained if the sources are intermittent and we are observing at a time when the disk contribution is off. While such an explaination is generally not comfortable, we propose here a model where the bulge-to-disk ratio is \emph{most of the time} as currently observed.

The temporary ``extinction'' of the disk component is only possible if the repetition timescale of the positron-generating events is larger than the annihilation timescale. Given the estimates of Sect.~\ref{sec:timescales}, this implies a repetition time larger than $10^5\,n_{0}^{-1}$~yr, ruling out a standard supernova origin or even more continuous sources related to the cosmic-ray interactions in the ISM. Gamma-ray bursts, on the other hand, are potentially interesting sources. From observations, one can derive that about one GRB could be \emph{observed} every $10^7$~yr in a galaxy like ours. Taking into account a beaming factor of 50--500 (Frail et al., 2001; Panaitescu \& Kumar, 2001), this would translate into an actual GRB rate of one every 20--200~kyr. If the jets are structured (Rossi et al., 2002), rather than homogeneous, 3--10 times more GRBs could be observed, leaving the actual rate around one every $10^5$--$10^6$~yr (Podsiadlowski et al., 2004). One the other hand, Wick et al. (2004) have argued for a higher frequency, around $10^{-4}\,\mathrm{yr}^{-1}$. Clearly, such a high frequency would rule our model out (if $n_{0} < 10\,\mathrm{cm}^{-3}$), while a rate of 1 GRB every $10^5\,n_{0}^{-1}$~yr (or more) would be favoured -- provided, of course, that GRBs can be a significant source of positrons. This point is now addressed.

\subsection{The origin of the positrons}

If the current e$^+$e$^-$ annihilation rate is not very different from its average value, the positron injection rate must be $\sim 1.3\,10^{43}\,\mathrm{s}^{-1}$, equivalent to $\sim 4\,10^{55}\Delta t_{5}$ positrons per GRB, where $\Delta t_{5}$ is the mean time between contributing events, in units of $10^5$~yr. To produce this by radioactive decay, the mass of $\beta^+$ nuclei (namely $^{56}$Co, via $^{56}$Ni) must be larger than $6\,\mathrm{M}_{\odot}\times \eta_{\mathrm{esc}}^{-1}\Delta t_{5}$, where we introduced the positron escape fraction, $\eta_{\mathrm{esc}} < 0.1$. This does not seem reasonable for a single GRB. However, GRBs also produce positrons with non radioactive origin. In a recent study, Furlanetto and Loeb (2002) recalled the two main processes producing positrons, inside and outside the fireball, and calculated the expected flux at 511~keV. Here, we briefly reanalyze these processes and discuss how the model should be modified in the case of GRBs associated with mini-starbursts in the central regions of the Galaxy.

The most natural source of GRB positrons is the fireball itself, which is believed to consist of a dense e$^+$e$^-$ pair plasma, initially at equilibrium with gamma-ray photons, with a total energy $E \equiv 10^{51}E_{51}$~erg. Assuming a typical Lorentz factor of 100, this amounts to $\sim 6\,10^{54}\,E_{51}$ positrons, which could be compatible with the required numbers if a significant fraction of these positrons could survive and escape out of the fireball. This, however, is not likely to be the case, because the recombination timescale inside the fireball, $t_{\mathrm{rec}} \sim [n_{\mathrm{\pm}}\left<\sigma v\right>_{\mathrm{rec}}]^{-1}$, is shorter than the dynamical timescale, $t_{\mathrm{dyn}}\sim R/c$. To be more precise, we note that in the early fireball, the dimensionless temperature of the pair plasma $\theta \equiv kT/m_{\mathrm{e}}c^2$ is larger than one and the thermal equilibrium maintains a large number of pairs. As the fireball expands, the temperature decreases as $R^{-1}$ (since the radiation pressure dominates) and when $\theta < 1$, the density of pairs is given by $n_{\pm}\simeq \sqrt{2/\pi^{3}}\alpha^3r_{\mathrm{e}}^{-3}\theta^{3/2}\exp(-1/\theta)$ (Shemi and Piran, 1990), where $\alpha$ is the fine structure constant and $r_{\mathrm{e}}$ is the classical radius of the electron. The \emph{freeze out} of positrons and electrons takes place when $t_{\mathrm{rec}}\sim t_{\mathrm{dyn}}$. For temperatures $\theta < 1$, one can write $\left<\sigma v\right>_{\mathrm{rec}}\simeq \pi r_{\mathrm{e}}^2 c$ (Svensson, 1982), and the freeze out condition reads:
\begin{equation}
\theta^{1/2}\exp(-1/\theta) = \frac{3}{8} \sqrt{\frac{\pi}{2}}\left(\frac{4}{11}\right)^{1/3}\frac{1}{\alpha^3\theta_{0}}\frac{r_{\mathrm{e}}}{R_{0}},
\label{eq:freezeOut}
\end{equation}
where $R_{0} \simeq 10$~km is the initial radius of the fireball, and $\theta_{0}\simeq 55$ its initial temperature, assuming a total energy of $10^{51}$~erg. Equation~\ref{eq:freezeOut} gives the freeze out temperature, $\theta_{\mathrm{fo}}\simeq 0.032$, which is very similar to the radiation temperature when the fireball becomes transparent to Thomson scattering in the case when the optical depth is dominated by the pairs (Shemi and Piran, 1990). We note, however, that pair freeze-out always occurs before transparency, because even if the fireball is devoid of external matter, it is slightly harder for a  positron to recombine than for a photon to interact (as $\left<\sigma v\right>_{\mathrm{rec}} < \sigma_{\mathrm{T}}c$ by a factor 3/8).

The total number of positrons which survive after recombination in the fireball is obtained as $N_{+} \simeq \frac{4}{3}\pi R_{\mathrm{fo}}^{3}n_{\pm}(\theta_{\mathrm{fo}})$, where $R_{\mathrm{fo}}\simeq R_{0}\theta_{0}/\theta_{\mathrm{fo}}(11/4)^{1/3}$ is the radius at the freeze-out time, and the factor $(11/4)^{1/3}$ accounts for the decrease in the number of degrees of freedom at the transition $\theta \la 1$. We obtain $N_{+}\simeq 5\,10^{43}$, which is many orders of magnitude less than the number of positrons required to account for the 511~keV data.

Efficient pair production is also expected to occur ahead of the relativistic fireball, as photons which are Compton backscattered by the ionized medium upstream interact with subsequent GRB photons via $\gamma\gamma$ pair-production interactions (Thompson and Madau, 2000; Dermer and B\"ottcher, 2000). In this case, the e$^+$e$^-$ pairs are produced with MeV energies, and assuming a rather conservative conversion efficiency, $\xi_{\mathrm{pair}} = 1$\% (Dermer and B\"ottcher, 2000), one obtains a total number of positrons $N_{+}\sim 6\,10^{54}\,E_{51}\,(\xi_{\mathrm{pair}}/0.01)$.

In a more detailed study of the spectral modifications of GRBs by pair precursors, M\'esz\'aros et al. (2001) obtained the positron yield for this mechanism at the transition between impulsive and wind-dominated evolution regimes: $N_{+} \sim 3\,10^{55}\,L_{50}^{9/10}t_{10}^{11/10}$, where $L_{50}$ is the GRB gamma-ray luminosity in units of $10^{50}\,\mathrm{erg~s}^{-1}$, and $t_{10}$ is the burst duration, as seen by a distant observer, normalized to 10~s. This is roughly proportional to the energy, $E_{\mathrm{51}} = L_{50}t_{10}$, and the obtained yield, $N_{+}\sim 3\,10^{55}E_{51}$, corresponds to a conversion efficiency of $\xi_{\mathrm{pair}}\sim 5$\%. This represents a very significant source of positrons which could in principle account for the annihilation line observed in the Galactic bulge. Combining this yield with the positron production rate required to explain the Galactic bulge data, one finds that GRBs could indeed account for the 511~keV emission if their repetition time in the Galactic bulge is roughly $\Delta t_{\mathrm{GRB}}\simeq 7.5\,10^4\,\mathrm{yr}\times E_{51}\xi_{0.05}$.

The GRB repetition time in a galaxy comparable to ours is quite uncertain, as recalled above, but values around $10^4$--$10^6$~yr seem reasonable. Therefore, the observed fluxes appear compatible with the above model, considering that $\ga 20$\% of the Galactic GRBs are expected to occur in the bulge. This derives from the current formation rate of massive stars in the bulge, namely $\sim 10$\% of that in the entire Galaxy (Figer, 2004), taking into account starburst episodes (see below) which increase the average value, and a discrimination in favour of massive binaries in the central regions of the Galaxy. This could be due to the effect of tidal forces allowing the survival of only the most massive molecular clouds, where GRB progenitors may form preferentially (Portegies Zwart et al., 2001, 2004). With a GRB energy of $1.3\,10^{51}$~erg (Bloom et al., 2003), one needs $\Delta t_{\mathrm{GRB}}\simeq 10^5$~yr in the bulge, and thus $\Delta t_{\mathrm{GRB}} \simeq 2\times 10^4$~yr in the whole Galaxy.

\subsection{Nuclear starbursts and positron injection}

In their study, Furlanetto and Loeb (2002) considered the 511~keV radiation induced by the annihilation of positrons associated with a GRB. Even if the positrons are produced outside the fireball, as argued above, and travel relativistically ahead of the jet, they will be swept up by the shock of the underlying supernova when the jet decelerates and grows sideways to form an isotropic remnant in spherical expansion. The positrons then mix with the ejecta on a timescale shorter than $10^3$~yr, for relevant values of the parameters. The positron annihilation time inside the remnant is always much larger than the age of the remnant, and therefore no significant annihilation occurs until the end of the Sedov-like phase when the shocked gas starts cooling radiatively and a dense shell forms at the shock front, at a time $t_{\mathrm{sf}}\simeq 3.6\,10^4\,E_{51}^{3/14}n_{0}^{-4/7}$~yr, where $n_{0}$ is the initial ambient density in cm$^{-3}$. Then the authors show that the positrons annihilate rapidly in the shell, and no signal persists after the a few $t_{\mathrm{sf}}$.

While such a scenario is appropriate for GRBs occurring in the disk of the Galaxy, we argue that the situation should be quite different in the case of GRBs associated with massive stars in the central regions. This is a simple consequence of the occurrence of mini starbursts driving large-scale bipolar winds in the Galactic center. This is now well established in nearby galaxies, and growing evidence shows that powerful mass ejections occur in the nucleus of our Galaxy, as resulting from the activity of giant star clusters involving masses of several $10^{6}\,\mathrm{M}_{\odot}$ and energies up to several $10^{53}$~erg (Bland-Hawthorn and Cohen, 2003, and refs. therein). The energy input, from individual stellar wind and multiple supernova explosions, should result in a complex, turbulent flow on scales as large as the Galactic bulge, very similarly to what occurs inside superbubbles, where the hot, low-density plasma sustained by the activity of an OB association blows a shell up to radii of half a kpc. Such starbursts are believed to occur repeatedly on timescales of a few Myr.

In a recent study of the energetic activity inside superbubbles, Parizot et al. (2004) showed that supernovae exploding in such an environment produce a shock which never becomes radiative, so that a dense shell never forms. This is due to the large sound velocity and low density, which implies that the shock becomes subsonic before being radiative. The case of a GRB exploding in the environment corresponding to a nuclear starburst in our Galaxy would be exactly the same: the associated supernova remnant where the positrons are trapped will not form a shell, and therefore the positrons will not be able to annihilate before the shock becomes subsonic and disappears. At that stage, the positrons are released in the ambient medium and diffusive transport allows them to fill the bulge, where they thermalize and eventually annihilate. Incidentally, referring to the line profiles calculated by Furlanetto and Loeb (2002), we note that the line width of positrons annihilating in a shell would be much too large to account for the data.

The mean time between nuclear starbursts of the order of a few $10^6$--$10^7$ yr, so that several GRBs are expected to occur in the same episode of activity. However, since the positron lifetime is always larger than $\Delta t_{\mathrm{GRB}}$ (if $n_{\mathrm{bulge}}\la 0.1\,\mathrm{cm}^{-3}$), a continuous emission is expected from the Galactic bulge, so that we do not need to refer to a particular recent event to account for the data. It is nevertheless possible that temporal variations of the order of a few in the gamma-ray flux accompany individual starbursts on Myr scales. In this respect, we note that the latest starburst occurred some 7~Myr ago, and could therefore contribute to an increased flux today.

On the other hand, the fact that the disk contribution should not exceed 0.5 -- 0.8 times that of the bulge is understood in our model as a natural consequence of the short annihilation timescale of the positrons from GRBs exploding in the disk. Most of the GRB progenitors are expected to be embedded in high a density medium, such as a dense molecular cloud around an OB association. For an explosion in the typical average density of a molecular cloud, $n = n_{30}\times 30\,\mathrm{cm}^{-3}$, we have $\tau_{\mathrm{ann}}\la 10^{4}\, n_{30}^{-4/7}$~yr (Furlanetto and Loeb, 2002), so that if $\Delta t_{\mathrm{GRB}} \ga 2\,10^{4}\,  n_{30}^{-4/7}$~yr in the whole Galaxy, the disk 511~keV line has a significant intensity for less than 50\% of the time. Even though massive stars usually explode in the low density medium filling superbubbles (formed by the activity of Wolf-Rayet stars and many supernova explosions in the OB association), it is expected that the GRB progenitors are the most massive of all the stars in the association, and therefore explode very early, before the density significantly drops. Contrary to the case of mini-starbursts in the central molecular zone of the Galaxy, where many GRBs explode during an extended period of time, not more than one GRB should explode in a given OB association, before the modification of the environment. It is nevertheless interesting to consider the case of a ``late'' GRB, exploding in an already formed superbubble. In this case, similarly to what occurs in the bulge, no shell will form around the associated supernova remnant for the positrons to readily annihilate. However, the superbubble itself is surrounded by a dense shell where the positrons will annihilate with a delay of $\sim 5\,10^5$~yr (Parizot et al., 2004). The corresponding annihilation signal could then be observed in one or a few superbubbles in the Galaxy. This is an interesting target for deeper surveys with INTEGRAL at large longitudes.

In this respect, it is worth noting that no complete survey has yet been made of the entire Galactic disk, with the same depth as in the bulge. In particular, the OSSE data do not allow one to exclude an emission from a particular superbubble, e.g. in the molecular ring at $\sim 30^\circ$. Thanks to its coded mask imaging technique, the INTEGRAL spectrometer will provide such a detailed map in the near future, and we suggest that the data should be searched for a 511 keV emission excess on a few degree scale (typically corresponding to the shell of a superbubble). We note finally that, in the context of our scenario, it is misleading to speak about a disk-to-bulge ratio, since the 511~keV emission is diffuse in the bulge, while localized in the disk, and thus there is no ``disk emission'' as such (from GRB positrons).

\section{Conclusion}

In this paper, we have shown that the phenomenology of the diffuse e$^{+}$e$^{-}$ annihilation emission in the Galactic bulge can be accounted for by positrons produced in $\gamma-\gamma$ interactions just ahead of the relativistic jet of GRBs. The flux level requires a mean time between GRBs occurring \emph{in the bulge} of $\Delta t_{\mathrm{GRB}} \simeq 7.5\,10^4\mathrm{yr}\,E_{51}\xi_{0.05}$. We have shown that the low disk-to-bulge emission ratio can be explained by differences in the environment of GRBs located in the disk and the Galactic bulge. In the latter case, GRBs are likely to occur during a starburst in a hot, low-density environment, where the associated supernova remnant dissipates before it can form a radiative shell. Therefore, the rapid annihilation of the positrons described by Furlanetto and Loeb (2002) does not occur in this case, and the positrons can diffuse in the ambient medium for $\sim 10^7$~years. This leaves them enough time to fill the bulge, if reasonable transport conditions apply. A crucial parameter of the model is the typical time between two GRBs in the Galaxy, which should be better constrained in the near future, notably with the help of the Swift satellite. If it turns out that $\Delta t_{\mathrm{GBR}}\ga 3\,10^4\,\mathrm{E}_{51}$~yr, the model will be excluded because the rate of positron injection in the bulge will be too small. On the other hand, if  $\Delta t_{\mathrm{GBR}}\la 10^4\,\mathrm{E}_{51}$~yr, the 511~keV emission from the positrons of the last few GRBs having exploded in the Galactic disk should not be extinguished, and in this case we anticipate a possible detection by INTEGRAL, in coincidence with one or a few superbubbles and/or OB associations in the Galaxy. It has been suggested recently that the supernova associated with remnant W49B was in fact a GRB, which occurred 3000~years ago (Ioka et al., 2004; Keohane et al. 2004). The possibility of detecting the associated e$^{+}$e$^{-}$ annihilation will be investigated in a forthcoming paper.

\begin{acknowledgements}
We thank Alexandre Marcowith and Michel Tagger for useful comments about positron transport, and the anonymous referee for his comments which helped us to clarify this paper.
\end{acknowledgements}


\begin{thebibliography}{}

\bibitem{} Armstrong, J.~W., Rickett, B.~J., \& Spangler, S.~R., 1995, ApJ, 443, 209

\bibitem{} Berezinskii, V.~S., Bulanov, S.~V., Dogiel, V.~A., \& Ptuskin, V.~S., 1990, Astrophysics of cosmic rays, ed. V.L. Ginzburg, North-Holland, Chapter 9

\bibitem{} Bertone, G., Kusenko, A., Palomares-Ruiz, S., Pascoli, S., \& Semikoz, D., 2004, astro-ph/0405005

\bibitem{} Bland-Hawthorn, J., \& Cohen, M., 2003, ApJ, 582, 246

\bibitem{Bloom+03} Bloom, J.~S., Frail, D.~A., \& Kulkarni, S.~R., 2003, ApJ, 594, 674

\bibitem{} Boehm, C., Hooper, D., Silk, J., Casse, M., \& Paul, J., 2004, Phys. Rev. Lett., 92, 101301

\bibitem{} Casse, F., Lemoine, M., \& Pelletier, G., 2002, Phys. Rev. D, 65, 023002 
\bibitem{} Cass{\' e}, M., Cordier, B., Paul, J., \& Schanne, S., 2004, ApJ, 602, L17

\bibitem{} Dermer, C.~D., \& B\"ottcher, M., 2000, ApJ, 534, L155

\bibitem{} Dermer, C.~D., B\"ottcher, M., \& Liang, E.~P., 2001, in Proc. of the Fourth INTEGRAL Workshop, ESA SP-459, 161

\bibitem{} Figer, D.F., 2002, Proc. IAU Symposium No. 212, eds. van der Hucht, K.A., Herrero, A., Esteban, C.

\bibitem{} Frail, D. A., Kulkarni, S. R., Sari, R., Djorgovski, S. G., Bloom, J. S., et al., 2001, ApJ, 562, L55

\bibitem{FurLoe02} Furlanetto, S.~R., \& Loeb, A., 2002, ApJ, 569, L91

\bibitem{} Ioka, K., Kobayashi, S., \& M{\' e}sz{\' a}ros, P., 2004, ApJ, 613, L17

\bibitem{} Keohane, J.~W., Reach, W.~T., Rho, J., \& Jarrett, T.~H., 2004, American 
Astronomical Society Meeting, 204

\bibitem{} M\'esz\'aros, P., Ramirez-Ruiz, E., \& Rees, M.~J., 2001, ApJ, 554, 660

\bibitem{} Panaitescu, A., \& Kumar, P.\ 2001, ApJ, 560, L49

\bibitem{} Parizot, E., 2004, in Proc. of the Cosmic Ray International Seminar, Nucl. Phys. B, Proc. Suppl. (in press)

\bibitem{} Parizot, E., Marcowith, A., van der Swaluw, E., Bykov, A., \& Tatischeff, V., 2004, A\&A, 424, 747

\bibitem{} Paul, J., 2004, Nucl. Instr. Meth. in Phys. Res. B, 221, 215

\bibitem{} Podsiadlowski, P., Mazzali, P.~A., Nomoto, K., Lazzati, D., \& Cappellaro, E., 2004, ApJ, 607, L17

\bibitem{} Portegies Zwart, S., Mc Millan, S., Hut, P., Makino, J., 2001, MNRAS, 321, 199

\bibitem{} Portegies Zwart, S., Mc Millan, S., Baumgardt, H., 2004, in ``The Formation and Evolution of Massive Young Star Clusters'', eds. Lamers, H.J.G.L.M., Nota, A. \& Smith, L.J., preprint in astro-ph/0403147

\bibitem{} Rossi, E., Lazzati, D., \& Rees, M.~J., 2002, MNRAS, 332, 945

\bibitem{} Schanne, S., Cass\'e, M., Cordier, B., \& Paul, J., 2004, Proc. of the
5th INTEGRAL Workshop, "The INTEGRAL Universe", Munich, Germany, preprint in
astro-ph/0404492

\bibitem{} Shemi, A., \& Piran, T., 1990, ApJ, 365, L55

\bibitem{} Spangler, S.~R., \& Gwinn, C.~R., 1990, ApJ, 353, L29

\bibitem{} Svensson, R., 1982, ApJ, 258, 321

\bibitem{} Thompson, C., \& Madau, P., 2000, ApJ, 538, 105

\bibitem{} Weidenspointner, G., Lonjou, V., Knoedlseder, J., Jean, P., Allain, M., et al., 2004, in Proc. of the 5th INTEGRAL Workshop, "The INTEGRAL Universe", Munich, Germany, preprint in
astro-ph/0406178

\bibitem{} Wick, S.~D., Dermer, C.~D., \& Atoyan, A., 2004, Astroparticle Physics, 21, 125

\bibitem{} Zhang, B. \& M{\' e}sz{\' a}ros, P., 2004, Int. Journal of Mod. Physics A, 18, in press
 
\end{thebibliography}
\end{document}